\documentclass{article}
\newcommand{\XX}[1]{\fbox{\footnotesize #1}}

\begin{document}

\title{
Further Results on Arithmetic Filters  for Geometric Predicates\thanks{
  \em 
  This work was partially supported by
  ESPRIT LTR GALIA and by
  the U.S. Army Research Office under grant DAAH04-96-1-0013.
  This work was done in part while O. Devillers was visiting Brown University.
} }

\author{
Olivier Devillers\thanks{INRIA, BP 93, 06902 Sophia Antipolis, France.
Olivier.Devillers@sophia.inria.fr}
\and Franco P. Preparata\thanks{Brown Univ.,
Dep. of Computer Science, Providence, RI 02912-1910 (USA). franco@cs.brown.edu}
}

\maketitle

\abstract{
An efficient technique to solve precision problems consists in
using exact computations.
For geometric predicates, using systematically expensive
exact computations can be avoided by the use of filters.
The predicate is first evaluated using rounding computations,
and an error estimation gives a certificate of the validity of the result.
In this note, we studies the statistical efficiency of filters
for cosphericity predicate with an assumption of regular distribution of
the points. We prove that the expected value of the polynomial
corresponding to the in sphere test is greater than $\epsilon$
with probability $O(\epsilon \log \frac{1}{\epsilon})$ improving
the results of a previous paper \cite{dp-papaf-98}.
}

\section{Introduction}
The assumption of real-number arithmetic, which is at the basis of conventional
geometric algorithms, has been seriously challenged in recent years,
since digital computers do not exhibit such capability.
Geometric algorithms involve the evaluation of predicates; to guarantee the 
structural correctness of the results, predicates must be evaluated
{\em exactly}.
 A geometric predicate  usually consists of evaluating
the sign of some algebraic expression.
In most cases, rounded computations yield a reliable result, but
sometimes rounded arithmetic introduces errors which may invalidate
the algorithms. Assuming error-free input data, the
rounded arithmetic may produce an incorrect result
only if the exact  absolute value of the algebraic expression is smaller
than some (small) $\varepsilon$, which  represents the largest error
that may arise in the evaluation of the expression. The threshold 
$\varepsilon$ depends on the  structure of the expression and on the
adopted computer arithmetic.
This is basically the philosophy behind the notion of {\em arithmetic filters},
whose function is to adjust the arithmetic overhead, so that no more
effort is expended than required by the test instance.

It is therefore of interest to estimate the frequency with which recourse to
arithmetic engines more powerful than standard platforms is necessary.
Such analysis must be carried out by making some {\em a priori} hypothesis on the
distribution of the input data, which are treated like random variables.
Since for our objectives only the absolute value of the algebraic
expressions is significant, hereafter "value" is to be intended as
"absolute value".

In a previous paper \cite{dp-papaf-98}, 
we have carried out such analysis for two crucial 
geometric predicates, the orientation test (which-side of a hyperplane) and
the insphere test (inside/ouside a hypersphere), on the hypotheses that the 
input points were uniformly distributed either in the unit ball
 ${\cal B}_{\delta}$ or in the unit cube ${\cal C}_{\delta}=[-1,1]^{\delta}$
in $\delta$-dimensional space.
Our results were that, for  a small value $V$, the probability that the
result of the orientation test is $<V$ is $\Theta(V)$ in all dimensions,
whereas for the  more complex insphere test we obtained bounds sublinear in
$V$ . Specifically,  we obtained $O(V^{2/3})$ in dimension 1 
(which is tight), $O(V^{1/2})$ in dimension
2, and $O(V^{1/2} \ln{V})$ in higher dimension. 

Later on , we discovered
 a discrepancy between these theoretical 
findings  for $\delta>1$ and the results of extensive simulations, 
which seemed to exhibit a linear behavior (see below).
This observation motivated a finer analysis, reported
in this note, whose conclusion is that for $\delta>1$
 and for $\delta +2$ points $p_1,p_2,\ldots,p_{\delta+2}$
uniformly chosen in the unit ball, the probability
that the value of the determinant, embodying the insphere test of
$p_{\delta+2}$ versus $p_1,p_2,\ldots,p_{\delta+1}$, is $<V$ is 
$O(V \ln (1/V)$, in closer agreement with the simulations.
The results extend to points uniformly chosen in a cube.
We also present an application of this analysis to  the three-dimensional
insphere test carried out with floating point arithmetic.

\section{ Analysis of the insphere test}
The  algebraic expression embodying the
predicate which tests if a point $p_{\delta+1}$ belongs
to the sphere  $S$ passing through points
$p_1p_2\ldots p_{\delta}$ and the origin, is the following determinant
\cite{dp-papaf-98}:

\[
\Delta_{\delta}
=
   \left| \begin{array}{cccc} x_{11}&x_{12}& \ldots &x_{11}^2+x_{12}^2+
\ldots+x_{1\delta}^2\\
 x_{21}&x_{22}& \ldots &x_{21}^2+x_{22}^2+\ldots+x_{2\delta}^2\\
\ldots\\
x_{\delta+1,1}&x_{\delta+1,2}& \ldots &x_{\delta+1,1}^2+x_{\delta+1,2}^2+\ldots+
x_{\delta+1,\delta}^2\end{array}\right|
\]

 As mentioned in the Introduction, in dimension 1 
the insphere test reduces to an in-interval test
and is only of moderate interest. Nevertheless, we have obtained
the following tight bound 
\cite{dp-papaf-98}
$Prob(|\Delta_1|\leq V) \leq \frac{17 \sqrt[3]{2}}{4} V^{\frac{2}{3}}
\simeq 5.36 V^{2/3}$

We now turn our attention to higher dimension, and let
$c=(\frac{c_1}{2},\ldots, \frac{c_{\delta}}{2})$ denote the center of the
sphere $S$. In the above determinant,
subtracting column $i$ times $c_i$ from the last column, enables
us to rewrite $\Delta_{\delta}$ as
\begin{eqnarray}
\Delta_{\delta}
&=&
   \left| \begin{array}{ccccc} x_{11}&x_{12}& \ldots &x_{1\delta}& 0\\
 x_{21}&x_{22}& \ldots &x_{2\delta}& 0\\
\ldots\\
x_{\delta,1}&x_{\delta,2}& \ldots &x_{\delta,\delta}& 0\\
x_{\delta+1,1}&x_{\delta+1,2}& \ldots &x_{\delta+1,\delta}&W\end{array}\right|\\
\vspace{.1in}
&=&
 |p_1p_2\ldots p_{\delta}|W
\end{eqnarray}
where
\[
W=(x_{\delta+1,1}^2+\ldots+x_{\delta+1,\delta}^2) -\sum_{i=1}^{\delta}
c_i x_{\delta+1,i}.
\]
Adding and subtracting $\sum \frac{c_i^2}{4}$ from the last expression
we obtain
\[
W= \sum_{i=1}^{\delta}(x_{\delta+1,i}-\frac{c_i}{2})^2 -\sum_{i=1}^{\delta}
(\frac{c_i}{2})^2.
\]
This expression can be more synthetically rewritten as
$W= |cp_{\delta+1}|^2-r^2$, i.e., $W$ is $ power(p_{\delta +1},S)$ 
of point $p_{\delta +1}$ with respect to the sphere $S$. Notice that
$ power(p_{\delta +1},S)$ is positive if $p_{\delta +1}$ is external to $S$
and negative if it's internal. Therefore random variable $\Delta_{\delta}$
is the product of the two random variables $|p_1p_2\ldots p_{\delta}|$ and
$ power(p_{\delta +1},S)$ ( of which, incidentally, $|p_1p_2\ldots p_{\delta}|$
has the form of a standard orientation test in dimension $\delta$).
Therefore to complete our analysis we must:
\begin{enumerate}
\item Analyze the statistical behavior of $|p_1p_2\ldots p_{\delta}|$;
\item Analyze the statistical behavior of $ power(p_{\delta +1},S)$;
\item Obtain a convenient upper bound to the product of two random variables.
\end{enumerate}
These tasks are the object of the next three subsections.
The main idea of the proof is to use the fact that $W=power(p_{\delta +1},S)$
does not depend actually on $p_1,p_2\ldots p_{\delta}$ but
only on their circumscribing sphere.

\subsection{Orientation test}
In \cite{dp-papaf-98} we have  shown that, given
$\delta$ points uniformly distributed in the unit ball  ${\cal B}_{\delta}$
in dimension $\delta$,
\[
Prob(|p_1,p_2\ldots p_{\delta}| \leq V) \leq \sigma_{\delta} V
\]
where $\sigma_{\delta}=\delta\frac{v_{\delta-1}^{\delta}}{v_{\delta}^{\delta-1}}$
and $v_{j}$ denotes the volume of the unit ball in dimension $j$.

In fact, these results can be  extended without any difficulty 
to the case in which the value of $|p_1,p_2\ldots p_{\delta}|$ is constrained
to an interval $[V,V+dV]$, by simply
changing in Equation (c) of \cite{dp-papaf-98} the integration bounds from
$\int_{a_{\delta}=0}^{\mbox{\footnotesize min}(V,a_{\delta-1})}$ to
$\int_{a_{\delta}=\mbox{\footnotesize min}(V,a_{\delta-1})}
     ^{\mbox{\footnotesize min}(V+dV,a_{\delta-1})}$.
This trivial modification readily yields
\begin{equation}
\label{det-eq}
Prob(V \leq |p_1,p_2\ldots p_{\delta}| \leq V+dV
          \; \mid \;  p_1,p_2\ldots p_{\delta}\in{\cal B}_{\delta}) 
\leq \sigma_{\delta} dV
\end{equation}

This result generalizes to the uniform
distribution in the unit cube ${\cal C}_{\delta}=[-1,1]^{\delta}$
as in \cite{dp-papaf-98}.

\begin{equation}
\label{det-cube-eq}
Prob(V \leq |p_1,p_2\ldots p_{\delta}| \leq V+dV
          \; \mid \; p_1,p_2\ldots p_{\delta}\in{\cal C}_{\delta}) 
\leq \psi_{\delta} dV
\end{equation}

where $\psi_{\delta}=\frac
{\delta v_{\delta} v_{\delta-1}^{\delta}\delta^{\frac{\delta(\delta-1)}{2}}}
{2^{\delta^2}}$.

\subsection{Power of a point with respect to a sphere}

Given  a sphere $S$, with center $c$ and radius $r$, we wish to compute
the probability for a random point $p$ to have a small (absolute value)
power with respect to $S$.

For a small value $V$ we observe that

\begin{small}
\[
\left(
|power(p,S)| = \left| |cp|^2 -r^2 \right| \leq V
\right)
\Longrightarrow
\left(
r-\frac{V}{2r} \leq \sqrt{r^2-V} \leq |cp| 
        \leq \sqrt{r^2+V} \leq r+\frac{V}{2r}
\right)
\]
\end{small}

Therefore the value of the power of $p$ with respect to $S$ is
smaller than $V$ if $p$ belongs to a spherical crown of $S$  of width
$\frac{V}{r}$. Clearly, the volume of such crown is given by the measure
(area) of $S$ multiplied by $\frac{V}{r}$, i.e., it is given by
$\delta v_{\delta} r^{\delta-1}\frac{V}{r}=\delta v_{\delta} r^{\delta-2} V$
( this holds in our hypothesis of small $V$).

Thus  $Prob(power(p,S) \leq V)$ is bounded
as follows:

\[
Prob(power(p,S) \leq V)
 \leq
\frac{\mbox{volume(crown}\cap\Omega)}{\mbox{volume}(\Omega)}
\]

The term $\mbox{volume(crown}\cap\Omega)$ is the product of $\frac{V}{r}$
by the area of $S\cap \Omega$.
At this point we  assume  $\Omega\subset {\cal C}_{\delta}$,
which is  obviously verified when $\Omega$ is either
${\cal B}_{\delta}$ or ${\cal C}_{\delta}$.
If $r<1$ we  bound  from above the volume of the crown by
$\delta v_{\delta} r^{\delta-2} V \leq \delta v_{\delta} V$.
If $r\geq 1$ we restrict ourselves to the portion of
the crown internal to ${\cal C}_{\delta}$ and obtain
$\mbox{area}(S\cap{\cal C}_{\delta})\frac{V}{r}\leq 
 \mbox{area}(S\cap{\cal C}_{\delta}) V
\leq \delta v_{\delta} V$.

In conclusion, we have

\begin{equation}
\label{power-eq}
Prob(power(p,S) \leq V \; | \; S \mbox{ given};\; p\in{\cal B}_{\delta})
\leq \frac{\delta v_{\delta}}{v_{\delta}}V = \delta V
\end{equation}

\begin{equation}
\label{power-cube-eq}
Prob(power(p,S) \leq V  \; | \; S \mbox{ given};\;  p\in{\cal C}_{\delta})
\leq \frac{\delta v_{\delta}}{2^{\delta}}V
\end{equation}

\subsection{Product of two  random variables}

To complete the analysis outlined above, 
we need a technical  result  concerning the probability of a product of
random variables.

\begin{figure} \begin{center}
\def\IPEfile{Hyperbole.ipe}\input{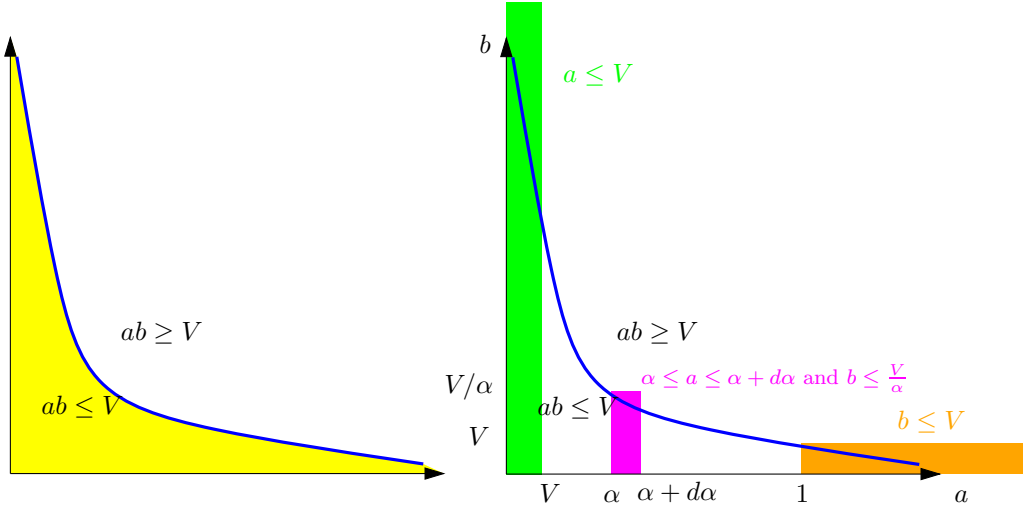}
\caption{Upper bounding event $ab\leq V$\label{Hyperbole}}
\end{center} \end{figure}

Let $a$ and $b$ be two  random variables such that the 
marginal probability of $a$ satisfies $Prob(V \leq a \leq V+dV)\leq AdV$
and the probability of $b$ conditional on $a$ satisfies
$Prob(b\leq V | a)\leq BV$, for some constants $A$ and $B$. 
Notice that our random
variables $|p_1,p_2\ldots p_{\delta}|$ and $power(p,S)$ fit the specifications
of $a$ and $b$, respectively.  
We shall bound from above the event $ab<V$
by a union of events of the kind
$\alpha \leq a \leq\alpha+d\alpha \mbox{ and } b\leq \frac{V}{\alpha}$,
as illustrated on Figure~\ref{Hyperbole}.

Thus we have

\begin{eqnarray} \nonumber
Prob(ab\leq V)
& \leq &
Prob(a\leq V) 
+ \int_{V}^{1} Prob(a=\alpha)Prob (b\leq \frac{V}{\alpha}| a=\alpha) d\alpha
+ Prob(b\leq V ) 
\\\nonumber& \leq &
(A+B)V + \int_{V}^{1} ABV \frac{d\alpha}{\alpha}
\\\label{product-eq}& \leq &
(A+B)V +  ABV \ln \frac{1}{V}
\end{eqnarray}

Notice that for $A$ and $B$ both $\geq 2$ and for $V \leq 1/e$, the first term 
is dominated by the second one.

\section{ Completing the analysis}

In this section, we  present the main
conclusion of this note. Recalling that
\[
\Delta_{\delta}= |p_1p_2\ldots p_{\delta}|.power(p_{\delta+1},{\mbox{sphere}(p_1p
_2\ldots p_{\delta})}),
\]
and the previous bounds, we obtain for the two domains:
\begin{equation}
Prob(\Delta_{\delta}\leq V | p_1\ldots p_{\delta+1} \in {\cal B}_{\delta} )
\leq \left(\sigma_{\delta} +\delta  \right)  V 
      + \sigma_{\delta} \delta V \ln \frac{1}{V}
\end{equation}

\begin{equation}
Prob(\Delta_{\delta}\leq V | p_1\ldots p_{\delta+1} \in {\cal C}_{\delta} )
\leq
 \frac{\delta v_{\delta}\psi_{\delta}}{2^{\delta}}V \ln \frac{1}{V}
 +
 \left(\psi_{\delta} +\frac{\delta v_{\delta}}{2^{\delta}} \right)  V
\end{equation}
which express a  bound nearly  linear in $V$ for the absolute value of the incircle
test for $\delta>1$.

For small values of $\delta$ we recall from \cite{dp-papaf-98} the (approximate) 
values of $v_{\delta}$, $\sigma_{\delta}$ and $\psi_{\delta}$:

\begin{center}
\begin{tabular}{|c|c|c|c|}
\hline 
$\delta$ & $v_{\delta}$        & $\sigma_{\delta}$       & $\psi_{\delta}$ \\\hline
1&2                            &  1                      & 1\\
2&$\pi$                        &$\frac{8}{\pi}\simeq 2.5$&$\pi \simeq 3.1$ \\
3&$\frac{4\pi}{3}\simeq 4.2$   & $\simeq 5.3$            & $\simeq 21$ \\
4&$\frac{\pi^2}{2}\simeq 4.9$  & $\simeq 10$             & $\simeq 380$ \\
5&$\frac{8\pi^2}{15}\simeq 5.3$& $\simeq 19$             & $\simeq 22.000$ \\
6&$\frac{\pi^3}{6}\simeq 5.2$  & $\simeq 35$             & $\simeq 4.500.000$ \\
\hline
\end{tabular}
\end{center}

\begin{eqnarray*}
Prob(\Delta_{2}\leq V | p_1\ldots p_{3} \in {\cal B}_{2} )
& \leq &
5.0  V\ln {\textstyle \frac{1}{V}} + 4.5    V \\
Prob(\Delta_{3}\leq V | p_1\ldots p_{4} \in {\cal B}_{3} )
& \leq &
 16 V\ln {\textstyle \frac{1}{V}} + 8    V \\
Prob(\Delta_{4}\leq V | p_1\ldots p_{5} \in {\cal B}_{4} )
& \leq &
40  V\ln {\textstyle \frac{1}{V}} +  14   V \\
Prob(\Delta_{5}\leq V | p_1\ldots p_{6} \in {\cal B}_{5} )
& \leq &
95  V\ln {\textstyle \frac{1}{V}} + 24    V \\
Prob(\Delta_{6}\leq V | p_1\ldots p_{7} \in {\cal B}_{6} )
& \leq &
207  V\ln {\textstyle \frac{1}{V}} +  40   V \\
\end{eqnarray*}
and \begin{eqnarray*}
Prob(\Delta_{2}\leq V | p_1\ldots p_{3} \in {\cal C}_{2} )
& \leq &
4.9  V\ln {\textstyle \frac{1}{V}} +  4.7   V \\
Prob(\Delta_{3}\leq V | p_1\ldots p_{4} \in {\cal C}_{3} )
& \leq &
32  V\ln {\textstyle \frac{1}{V}} + 22    V \\
Prob(\Delta_{4}\leq V | p_1\ldots p_{5} \in {\cal C}_{4} )
& \leq &
468  V\ln {\textstyle \frac{1}{V}} + 381    V \\
Prob(\Delta_{5}\leq V | p_1\ldots p_{6} \in {\cal C}_{5} )
& \leq &
18000  V\ln {\textstyle \frac{1}{V}} +  22000   V \\
Prob(\Delta_{6}\leq V | p_1\ldots p_{7} \in {\cal C}_{6} )
& \leq &
2.200.000  V\ln {\textstyle \frac{1}{V}} +  4.500.000   V \\
\end{eqnarray*}

\begin{figure} \begin{center}
\def\IPEfile{Experiment.ipe}\input{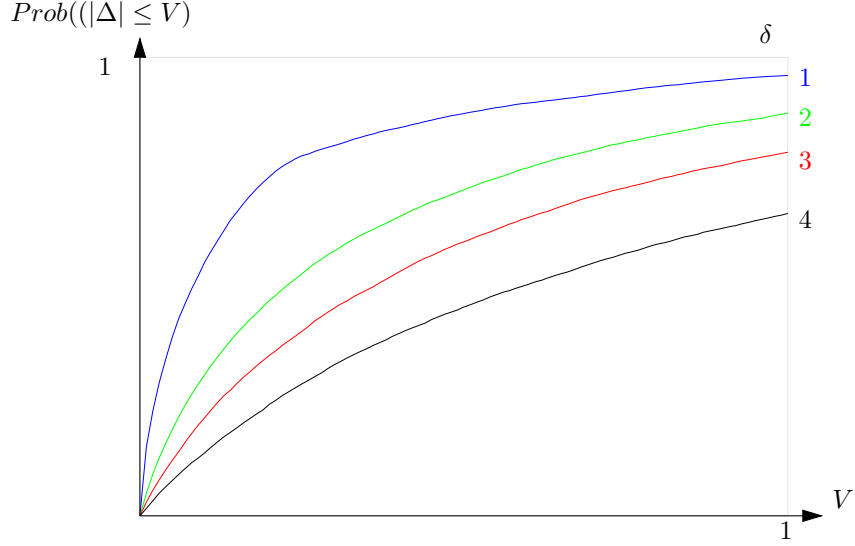}
\caption{\label{Experiment} Experimental results on random incircle tests}
\end{center} \end{figure}

These analytical results can be compared with the experimental results
mentioned earlier. The latter have been obtained using random point
selection in ${\cal B}_{\delta}$, and are shown in Figure \ref{Experiment}.
They confirm the sublinear behavior for $\delta=1$ and a basically
linear behavior for $\delta\geq 2$ near $V=0$. However, the constants reported
above are far from tight when the dimension increases, which is a clear
byproduct of the technique of proof used in \cite{dp-papaf-98}
to bound $\psi_{\delta}$. 

\section{Example: \normalsize 3D insphere test with double precision floating point arithmetic}

We now consider a practical implementation of the insphere test in three dimensions.
The corresponding expression is
given below. We assume that entries
(point coordinates) are floating point numbers in the range $[-1,1]$
and that they are stored as double precision numbers with a 53-bit mantissa.
We assume that the computation complies with the IEEE 754 norm.

We first detail the formula for the insphere test:
\[
\left|
\begin{array}{cccc}
x_1 & y_1 & z_1 & x_1^2+y_1^2+z_1^2 \\
x_2 & y_2 & z_2 & x_2^2+y_2^2+z_2^2 \\
x_3 & y_3 & z_3 & x_3^2+y_3^2+z_3^2 \\
x_4 & y_4 & z_4 & x_4^2+y_4^2+z_4^2 
\end{array} 
\right|
=
- (x_1^2+y_1^2+z_1^2) 
\left|\begin{array}{ccc}
x_2 & y_2 & z_2 \\
x_3 & y_3 & z_3 \\
x_4 & y_4 & z_4 
\end{array} \right|
+ (x_2^2+y_2^2+z_2^2)
\left|\begin{array}{ccc}
x_1 & y_1 & z_1 \\
x_3 & y_3 & z_3 \\
x_4 & y_4 & z_4 
\end{array} \right|
\] \[ \hspace*{5cm}
- (x_3^2+y_3^2+z_3^2)
\left|\begin{array}{ccc}
x_1 & y_1 & z_1 \\
x_2 & y_2 & z_2 \\
x_4 & y_4 & z_4 
\end{array} \right|
+ (x_4^2+y_4^2+z_4^2)
\left|\begin{array}{ccc}
x_1 & y_1 & z_1 \\
x_2 & y_2 & z_2 \\
x_3 & y_3 & z_3 
\end{array} \right|
\]
\[
\left|\begin{array}{ccc}
x_1 & y_1 & z_1 \\
x_2 & y_2 & z_2 \\
x_3 & y_3 & z_3 
\end{array} \right|
=
x_1 (y_2 z_3-y_3z_2) - x_2 (y_1 z_3 - y_3 z_1) + x_3 (y_1 z_2 - y_2 z_1)
\]

We now estimate the maximum {\em a priori} round-off error
using the following standard rules:
$\mbox{error}(x+y)\leq\mbox{error}(x)+\mbox{error}(y)+(x+y)2^{-54}$ and
$\mbox{error}(xy)\leq x\times \mbox{error}(y)+y \times \mbox{error}(x)+xy.2^{-54}$.
Each computation is analyzed in terms of  the  elementary operations of
addition/subtraction or multiplication.


\smallskip

\noindent
\begin{tabular}{|c|c|r|r|l|}
\hline
ref & description           & typical expression      & upper bound & error bound \\
\hline
\XX{1}& entry               & $x_1$                   & 1  & $2^{-54}$ \\
\XX{2}&\XX{1} $\times$ \XX{1} & $y_2 z_3$               & 1  & $3.2^{-54}$ \\
\XX{3}&\XX{2} +        \XX{2} & $y_2 z_3-y_3z_2$        & 2  & $2.3.2^{-54}+2.2^{-54
} = 2^{-51}$\\
\XX{4}&\XX{1} $\times$ \XX{3} & $x_1 (y_2 z_3-y_3z_2)$  & 2  & $2^{-51}+2.2^{-54} =
5.2^{-53}$\\
\XX{5}&\XX{4} +        \XX{4} &                         & 4  & $2.5.2^{-53}+4.2^{-54
} = 3.2^{-51}$\\
\XX{6}&\XX{5} +        \XX{4} & $\left|\begin{array}{ccc}
x_1 & y_1 & z_1 \\
x_2 & y_2 & z_2 \\
x_3 & y_3 & z_3
\end{array} \right|$                                  & 6  & $3.2^{-51}+5.2^{-53}+6.
2^{-54} = 5.2^{-51}$\\
\XX{7}&\XX{2} +        \XX{3} & $x_1^2+y_1^2+z_1^2$     & 3  & $3.2^{-54}+2^{-51}+3.
2^{-54} = 7.2^{-53}$\\
\XX{8}&\XX{6} $\times$ \XX{7} &                         & 18 & $6.7.2^{-53}+3.5.2^{-
51}+18.2^{-54} = 111.2^{-53}$\\
\XX{9}&\XX{8} +        \XX{8} &                         & 36 & $2.111.2^{-53}+36.2^{
-54} = 120.2^{-52}$\\
\XX{10}&\XX{9} +       \XX{9} & incircle test           & 72 & $2.120.2^{-52}+72.2^{
-54} = 129.2^{-51}\simeq 2^{-44}$\\
\hline \end{tabular}
\smallskip

If the points are uniformly distributed in the unit cube and  snap-rounded to
the nearest representable point, then the above calculations show  that if the
insphere test gives a result larger than $129\;2^{-51}$ (in absolute value),
then its sign is reliable.

For simple precision numbers with 24
 bits of mantissa, an analogous statement can be made for results larger than
$129\;2^{-22}\simeq 2^{-15}$.

These results enable us to estimate the probability of failure of such filter,
i.e., $prob({\rm failure}) \leq 32 (V)\ln\frac{1}{V}+22 (V)$
with  $V = 129.2^{-51}$ or $V = 129.2^{-22}$ for the two cases.

\begin{quote}
{\bf Claim:} \em
If the absolute value of the insphere test in three dimensions for
points in the unit cube computed with $53$ (resp. $24$) bit arithmetic is
larger than $129\;2^{-51}\leq 6\;10^{-14}$ (resp. $129\;2^{-22}\simeq 3\;10^{-5}$)
then the sign is reliable.
The probability of failure of the certifier is less than $6 \; 10^{-11}$
(resp. $0.011$).
\end{quote}
.

\end{document}